\documentclass{gGAF2e}

\usepackage[normalem]{ulem}
\usepackage{amsfonts,amsmath,amssymb}
\usepackage{environ}
\usepackage{graphicx}
\usepackage[breaklinks=true,colorlinks=true,allcolors=blue]{hyperref}
\usepackage{breakurl}
\usepackage[leftcaption]{sidecap}
\usepackage{epsfig}
\usepackage{psfrag}
\newcommand{\mypsfrag}[2]{\psfrag{#1}{\small{#2}}}
\usepackage{tabularx}
\usepackage{ragged2e}
\newcolumntype{Y}{>{\RaggedRight\arraybackslash}X}
\usepackage{booktabs}
\usepackage{xspace}
\usepackage[UKenglish]{babel}
\usepackage{placeins}

\renewcommand{\vec}[1]{\mbox{\boldmath $ #1$}}
\renewcommand{\v}{\vec}

\newcommand{\B}{\vec B}
\renewcommand{\u}{\vec u}

\newcommand{\ru}{\hat{\vec r}}
\renewcommand{\r}{\vec r}
\renewcommand{\k}{\hat{\vec k}}

\newcommand{\rhobar}{\bar{\rho}}
\newcommand{\Tbar}{\bar{T}}
\newcommand{\Pbar}{\bar{P}}

\newcommand{\Nrho}{N_\rho}
\newcommand{\pol}{v}
\newcommand{\tor}{w}

\newcommand{\R}{\mathrm{R}}
\renewcommand{\Pr}{\mathrm{Pr}}
\newcommand{\Pm}{\mathrm{Pm}}

\newcommand{\Z}{{\mathrm{Z}}}
\providecommand{\Emps}{\ensuremath{\overline{E}_p^s}\xspace}

\providecommand{\Emts}{\ensuremath{\overline{E}_t^s}\xspace}

\usepackage{tabularx}
\usepackage{ragged2e}
\newcolumntype{Y}{>{\RaggedRight\arraybackslash}X}
\usepackage{booktabs}

\allowdisplaybreaks

\begin{document}

\jvol{xx} \jnum{xx} \jyear{2017} \jmonth{}

\markboth{Simitev \& Busse}{Baroclinically-driven flows and dynamos}

\title{Baroclinically-driven flows and dynamo action  \\ in rotating
  spherical fluid shells} 

\author{RADOSTIN D. SIMITEV${\dag}$$^{\ast}$\thanks{$^\ast$Corresponding author. Email: Radostin.Simitev@glasgow.ac.uk
\vspace{6pt}} and FRIEDRICH H. BUSSE${\ddag}$\\\vspace{6pt}
  ${\dag}$School of Mathematics and Statistics, University of
  Glasgow, Glasgow G12 8SQ, UK\\ ${\ddag}$Institute of Physics,
  University of Bayreuth, 95440 Bayreuth,  Germany\\\vspace{6pt}\received{\today} }
\maketitle

\begin{abstract}
The dynamics of stably stratified stellar radiative zones is of
considerable interest due to the availability of increasingly detailed
observations of Solar and stellar interiors.
This article reports the first non-axisymmetric and time-dependent
simulations of flows of anelastic fluids driven by baroclinic
torques in stably stratified rotating spherical shells -- a system
serving as an elemental model of a stellar radiative zone. With
increasing baroclinicity a sequence of bifurcations from simpler to
more complex flows is found in which some of the available symmetries
of the problem are broken subsequently. The poloidal component of the
flow grows relative to the dominant  toroidal  component with
increasing baroclinicity. The possibility of magnetic field generation
thus arises and this paper proceeds to provide some indications for
self-sustained dynamo action in baroclinically-driven flows. We
speculate that magnetic fields in stably stratified stellar interiors
are thus not necessarily of fossil origin as it is often assumed.
\end{abstract}

\section{Introduction}
Stellar radiation zones are typically supposed to be motionless in standard
models of stellar structure and evolution \citep{Kippenhahn2012}, but
this assumption is poorly justified
\citep{Pinsonneault1997,Lebreton2000}. Indeed, increasingly detailed
evidence is emerging, e.g.~from helio- and asteroseismology, of
dynamical processes such as differential rotation, meridional
circulation, turbulence, and internal waves in radiative zones 
\citep{TurckChieze2008,Thompson2003,Gizon2010,Chaplin2013,Aerts2010}. These 
transport processes are important in mixing of angular momentum,
evolution of chemical abundances, and magnetic field sustenance
\citep{Miesch2009,Mathis2013}. The problem of fluid motions in
radiative zones is thus of wide significance.   

Historically, this problem has  been of much interest ever since
it became apparent that rotating stellar interiors cannot be in a static
equilibrium \citep{Zeipel}, a statement known as \emph{von Zeipel's
paradox}.  Two different forms of flow have been hypothesized,
namely (a) steady low-amplitude meridional circulations 
\citep{Vogt1925,Eddington1925,Eddington1925a}, and (b) particular
forms of strong differential rotation with minimal meridional circulations
\citep{Schwarzschild1947,Roxburgh1964}. It was demonstrated
that hypothesis (a) is not an acceptable solution to von Zeipel's
paradox \citep{Busse1981,Busse1982}.  Hypothesis (b) has since gained
support with self-consistent quasi-one-dimensional solutions first
derived by \cite{Zahn1992} and their baroclinic instabilities studied
\citep[e.g.~by][]{Spruit1984,Caleo2016}. 
In recent years a number of numerical studies of axisymmetric and
steady baroclinically-driven flows of finite amplitudes in rotating,
stably stratified spherical shells have been published 
\citep{Garaud2002,Rieutord2006a,EspinosaLara2013Selfconsistent,Hypolite2014Dynamics,Rieutord2014Dynamics}.  
In \citep{Garaud2002} the dynamics of the radiative zone of the Sun 
driven by the differential rotation of the  convective zone is
investigated, but the flow is essentially driven by the boundary
conditions and baroclinicity is not the main driving force. 
The cited papers by Rieutord and coworkers  offer perhaps the most
detailed studies of the problem to date. However, the analysis is
strongly restricted in two respects: first, two-dimensional and steady
axisymmetric flows are studied, and second, the Boussinesq
approximation is used which does not account for the strong density
variations typical for stably stratified regions of stars.

In this paper we present a model of baroclinic flows in rotating
spherical fluid shells. The model is based on the anelastic
approximation \citep{Gough1969,BraginskyRoberts1995,LantzFan1999} of the
fully-compressible fluid equations which is widely adopted for
numerical simulation of convection in Solar and stellar interiors
\citep{JonesBench}. {While this approximation is strictly only
valid for a system close to an adiabatic state it, nevertheless,
relaxes the assumptions of the Boussinesq approximation that has been
used previously.} 
{In the basic state the density is stably stratified in the radial
direction and axisymmetric shear driven motions are realized.}
Starting with this basic axisymmetric baroclinic state we explore the
onset of non-axisymmetric and time dependent states and investigate
their nonlinear properties. Further, we consider the possibility of
magnetic field generation by these flows.
Although a prime motivation for our study is to understand the motions
in stellar radiative zones, we have not strived to adjust the
values of the dimensionless parameters in our model to physically
realistic ones.  
{This is not possible in any case since the Reynolds number
  of flows in stars is huge and turbulent mixing occurs. Only the
  large scale flows can be simulated numerically while the influence
  of turbulence is taken into account through the use of eddy
  diffusivities in the equations of motion. For details of this
  approach we refer to the work of \cite{Miesch2009} and references therein.
Using this approach we shall
overcome the restrictions of axisymmetric steady flows which are
likely to be unstable.  As we shall show non-axisymmetric and time
dependent flows must be expected instead with properties that could
give rise to dynamo action.} 

\section{Mathematical model}
We consider a perfect gas confined to a spherical shell
{rotating} with a fixed angular velocity $\varOmega \k$ and with
a positive entropy contrast $\Delta S$ imposed between its outer
and inner surfaces at radii $r_o$ and $r_i$, respectively.  
{We assume a gravity field proportional to $g/r^2$. To
justify this choice consider the Sun, the star with the best-known
physical properties. The Solar density drops from 150 g/cm$^3$ at the
centre to 20 g/cm$^3$ at the core-radiative zone boundary (at $0.25
R_\odot$) to only 0.2 at the tachocline (at $0.7 R_\odot$). A crude
piecewise linear interpolation shows that most of the mass is
concentrated within the core. In this setting} a hydrostatic
polytropic reference state 
exists with profiles of density    $\rhobar = \rho_c\zeta^n$,
temperature $\Tbar=T_c\zeta$ and pressure $\Pbar = P_c \zeta^{n+1}$,
where $\zeta= c_0+c_1/r$ and $c_0=(2\zeta_o-\eta-1)/(1-\eta)$,
$c_1=(1+\eta)(1-\zeta_o)/(1-\eta)^2$, $\zeta_o=(\eta+1)/(\eta
\exp(\Nrho/n)+1)$, see \citep{JonesBench}.
The parameters $\rho_c$, $P_c$ and $T_c$ are reference values of
density, pressure and temperature at mid-shell. The gas polytropic
index $n$, the density scale height $N_\rho$ and the shell radius
ratio $\eta$ are defined further below. 
Following \cite{Rieutord2006a} we neglect the
distortion of the isopycnals caused by the centrifugal force. The
governing anelastic equations of continuity, momentum and energy
(entropy) take the \hfill\\
\newpage
\noindent
form 
\begin{subequations}
\label{govmod2}
\begin{align}
\label{govmod2.01}
{\bm \nabla}{\bm \cdot}\rhobar\u =& 0, 
\\
\upartial_t \u + ({\bm \nabla}\times\u)\times\u = &-{\bm \nabla}\varPi
-\tau(\k\times\u)+\frac{\R}{\Pr}\frac{{S}}{r^{2}}\ru + \v F_\nu \nonumber \\
\phantom{=}&\qquad\quad  - \Z (\overline{S}+S)\;  \k\times(\r\times\k), \label{govmod2.02}\\
\label{govmod2.03}
\upartial_t S + \u{\bm \cdot}{\bm \nabla}{(\overline{S}+S)}
 =& \frac{1}{\Pr \rhobar\Tbar} {\bm \nabla}{\bm \cdot}\rhobar\Tbar {\bm \nabla} S
  + {\frac{c_1 \Pr}{\R \Tbar}Q_\nu},
\end{align}
\end{subequations}
where
$S$ is the deviation from the background entropy profile
$\overline{S}=(\zeta(r)^{-n}-\zeta_o^{-n})/(\zeta_o^{-n}-\zeta_i^{-n})$ and
$\u$ is the velocity vector, ${\bm \nabla} \varPi$ includes all terms that can
be written as gradients,  and $\r=r \ru$ is the position vector with
respect to the center of the sphere \citep{JonesBench,Simitev2015}.  The viscous force $\v F_\nu =
(\rho_c/\rhobar){\bm \nabla}{\bm \cdot}\v {\hat S}$ and the viscous
heating $Q_\nu=\v{\hat S}{\bm :}\v e$
are defined in terms of the deviatoric stress tensor  
$\hat S_{ij}=2\rhobar(e_{ij}-e_{kk}\delta_{ij}/3)$ with
$e_{ij}=(\upartial_i u_j +\upartial_j u_i)/2$,
where double-dots (\textbf{:}) denote a Frobenius inner product, and
$\nu$ is a constant viscosity. The governing equations \eqref{govmod2}
have been non-dimensionalized with the shell thickness $d=r_o-r_i$ as unit of
length, $d^2/\nu$ as unit of time, and $\Delta S$, $\rho_c$ and
$T_c$ as units of entropy, density and temperature, respectively. 
The system is then characterized by seven dimensionless parameters:
the radius ratio $\eta=r_i/r_o$, 
the polytropic index of the gas $n$,
the density scale number $N_\rho=\ln\big(\rhobar(r_i)/\rhobar(r_o)\big)$,
the Prandtl number $\Pr={\nu}/{\kappa}$,
{the Rayleigh number $\R=-{gd^3\Delta S}/({\nu\kappa c_p})$,} 
the baroclinicity parameter 
$\Z={\varOmega^2 d^4 \Delta S}/({\nu^2 c_p})$,
and the Coriolis number $\tau = {2\varOmega d^2}/{\nu}$,
where $\kappa$ is a constant entropy diffusivity and $c_p$ is the
specific heat at constant pressure. {Note, that the Rayleigh
number assumes negative values in the present problem since the
basic entropy gradient is reversed with respect to the case of
buoyancy driven convection.} 
The poloidal-toroidal decomposition
\begin{gather*}
\rhobar \vec u = {\bm \nabla} \times ( {\bm \nabla} \times \ru r\pol) + {\bm \nabla} \times
\ru r^2 \tor
\end{gather*}
is used to enforce the solenoidality of the mass flux $\rhobar \u$.
This has the further advantages that the pressure gradient is 
eliminated and scalar equations for the poloidal and the toroidal
scalar fields, $v$ and $w$, are obtained by taking $\ru{\bm \cdot}{\bm \nabla}{\times{\bm \nabla}\times}$
and $\ru{\bm \cdot}{\bm \nabla}\times$ of equation \eqref{govmod2.02}. Except for the term
with the baroclinicity parameter $\Z$ the resulting equations are
identical to those described by \cite{Simitev2015}.    
{Assuming that entropy fluctuations are damped by convection in the
region above $r=r_o$ we choose the boundary condition}
\begin{subequations}
\label{BC}
\begin{align}
  S&=0 \;\; &\text{at } \qquad r&=\left\{\begin{array}{l}
 r_i\equiv\eta/(1-\eta)\,,\\[0.1em]
 r_o\equiv1/(1-\eta)\,,
  \end{array}\right.  \qquad
\intertext{while the inner and the outer boundaries of the shell are assumed
stress-free and impenetrable for the flow}
\pol = 0, \quad
\upartial_r^2 \pol = \frac{\rhobar'}{\rhobar r}\upartial_{r}{} (r\pol), \quad
\upartial_r (r \tor)& = \frac{\rhobar'}{\rhobar} \tor \;
&\text{at } \qquad r&=\left\{\begin{array}{l}
 r_i\,,\\[0.1em]
 r_o\,.
  \end{array}\right.
\end{align}
\end{subequations}

\section{Numerical solution}
For the numerical solution of problem (\ref{govmod2}--\ref{BC}) 
a pseudo-spectral method described by \cite{Tilgner1999} was employed.
A code developed and used by us for a number of years
\citep{Busse2003a,Simitev2011b,Simitev2015} and extensively
benchmarked for accuracy \citep{Simitev2015,Marti2014,Matsui2016} 
was adapted. 
Adequate numerical resolution for the runs has been chosen as described in
\citep{Simitev2015}.        
To analyse the properties of the solutions we
decompose the kinetic energy density into poloidal and toroidal
components and further into mean (axisymmetric) and fluctuating
(nonaxisymmetric) components and into equatorially-symmetric and
equatorially-antisymmetric components, 
\begin{align}
\label{Engs}
\overline E_p = \overline E_p^s +\overline E_p^a &=\langle \big({\bm \nabla} \times ( {\bm \nabla} \bar \pol \times \vec r)
\big)^2/(2\rhobar)  \rangle, \nonumber\\
\overline E_t = \overline E_t^s +\overline E_t^a& = \langle \big({\bm \nabla} r \bar\tor \times \vec r \big)^2
/(2\rhobar)  \rangle, \nonumber \\
\widetilde E_p = \widetilde E_p^s +\widetilde E_p^a&= \langle \big({\bm \nabla} \times ( {\bm \nabla} \check \pol
\times \vec r)  \big)^2/(2\rhobar) \rangle, \nonumber\\
\widetilde E_t = \widetilde E_t^s +\widetilde E_t^a&= \langle \big( {\bm \nabla} r \check \tor \times\vec r \big)^2
/(2\rhobar) \rangle, \nonumber
\end{align}
where  angular brackets $\langle\,\, \rangle$ denote averages 
over the volume of the spherical shell. 

\section{{Parameter values and initial conditions}}
In the simulations presented here fixed values are used for all
governing parameters except for the baroclinicity $\Z$ 
namely $\eta=0.3$, $n=2$, $N_\rho=2$, $\Pr=0.1$,
  $\tau=300$ and $\R=-5 \times 10^4$.
The value for the shell thickness represents the presence of a stellar
core {geometrically similar to that of the Sun.} 
The values for $n$ and $N_\rho$ {are not far removed from
estimates $n=1.5$ and $\Nrho = 4.6052$ for the solar radiative zone.}

{The strength of the baroclinic forcing, measured by $\Z$, is
limited from above so that
$$
\Z < (1-\eta)^3\; |\R|/\Pr.
$$
This restriction guarantees that the apparent gravity does not point
outward such that the model excludes standard thermal convection
instabilities.} 

Unresolved subgrid-scales in convective envelopes are typically
modelled by the assumption of approximately equal turbulent eddy
diffusivities and the choice of $\Pr=1$ is often made in the
literature \citep{Miesch2015}. However, in the presence of the minute
Prandtl number values based on molecular diffusivities the eddy
diffusivities are not likely to yield an effective Prandtl number of
the order unity. 
Furthermore, in a stably stratified system turbulence is expected to
be anisotropic \citep{Zahn1992}. With this in mind we have chosen
$\Pr=0.1$ in our study. 
\begin{figure}
\begin{center}
\epsfig{file=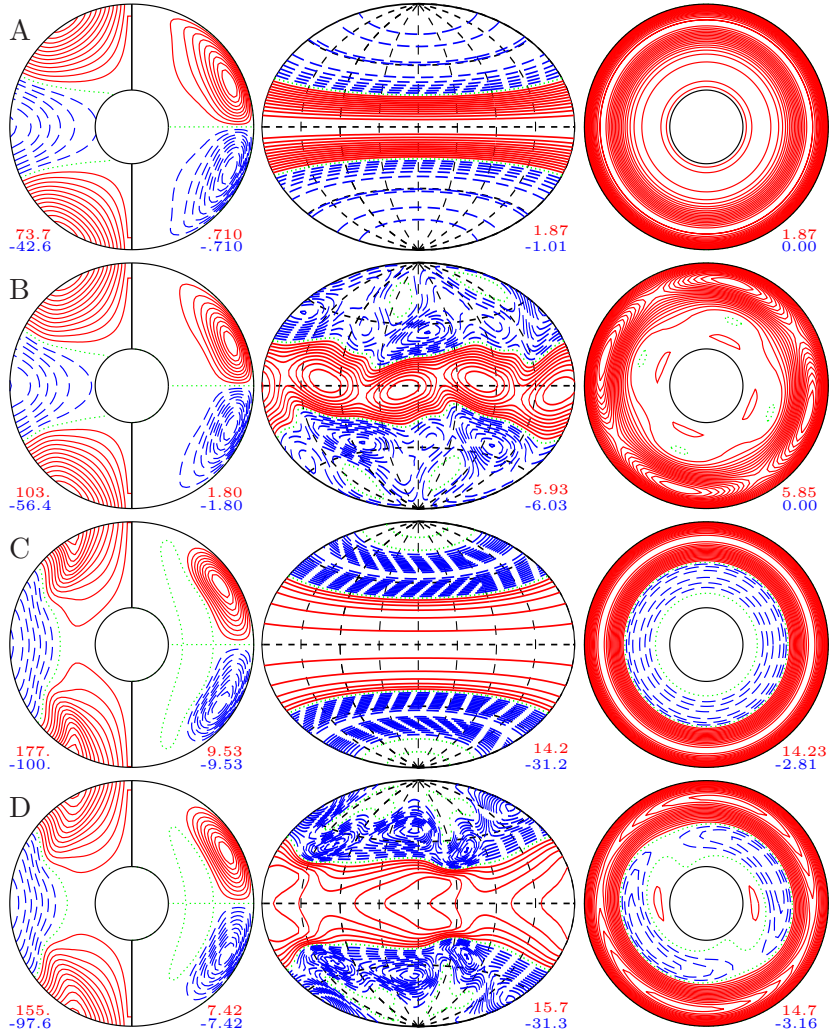,width=0.7\textwidth,clip=}
\end{center}
\caption{
  Flow structures with increasing baroclinicity 
  $\Z \times 10^{-4} = 7$, $10$, $14$ and $14$ from
  top to bottom. Bistability occurs at $\Z \times 10^{-4} = 14$. The values of the other parameters are $\eta=0.3$,
  $n=2$, $N_\rho=2$, $\Pr=0.1$, $\tau=300$ and $\R=-5 \times 10^4$. The first  plot in each row shows isocontours of
  $\overline{u}_\varphi$ (left half) and streamlines
  $r\sin\theta(\upartial_\theta \overline{\pol})=$ const. (right half)
  in the meridional plane. The second plot shows
  isocontours of $u_r$ at $r=r_i+0.7$. The third plot shows
  isocontours of $u_r$ in the equatorial plane. {The minimal
  and the maximal values of each field are listed under the
  corresponding plot; The isocontours are equidistant with positive
  isocontours shown by solid lines,   negative isocontours shown by
  broken lines and the zeroth isocontour shown by a dotted line in each plot.}
(Colour online)
}
\label{fig:010}
\end{figure}

The value $\tau = 300$ offers a good compromise in which the
effect of rotation is strong enough to govern the dynamics of the
system, but not too strong to cause a significant increase of the
computational expenses; similar values are used e.g.~by
\cite{Simitev2015} and in cases F1--4 of \citep{Kapyla2016}.
A negative $\R=-5 \times 10^4$ is assumed to model a convectively 
stable situation. 

Initial conditions of no fluid motion are used at vanishingly small
values of $\Z$, while at finite values of $\Z$ the closest equilibrated neighbouring case is
used as initial condition to help convergence and reduce
transients. To ensure that transient effects are eliminated from the
sequence presented below, all solutions have been continued for at
least 15 time units.  
\begin{figure}
\begin{center}
\mypsfrag{Ex}{$E_x$}
\mypsfrag{z}{\hspace{0mm}$\Z$}
\mypsfrag{ratio}{ratios}
\epsfig{file=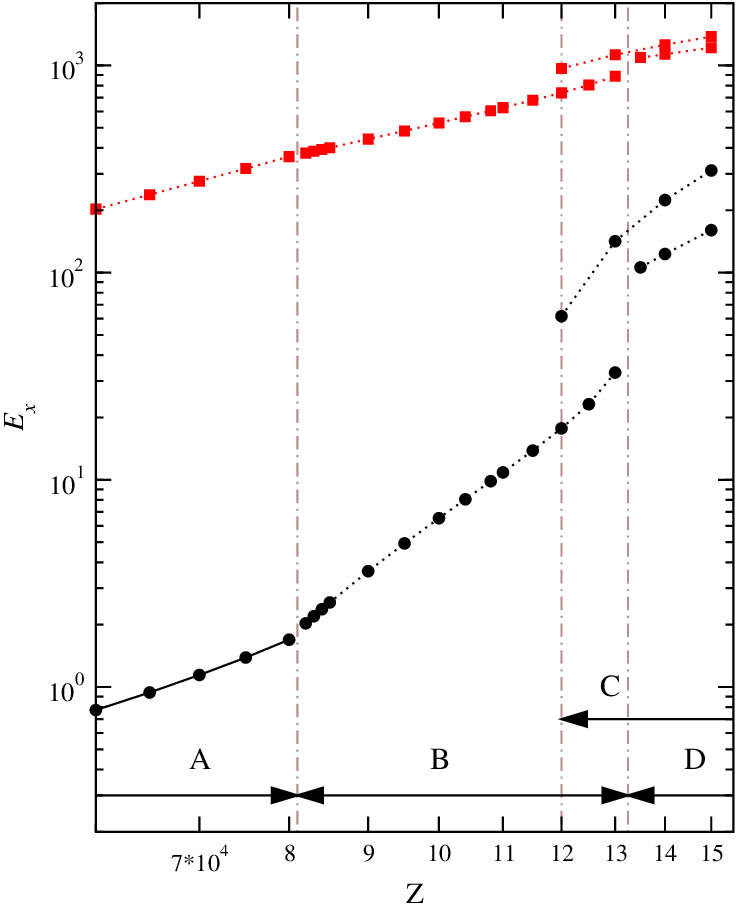,width=0.48\textwidth,clip=}
\epsfig{file=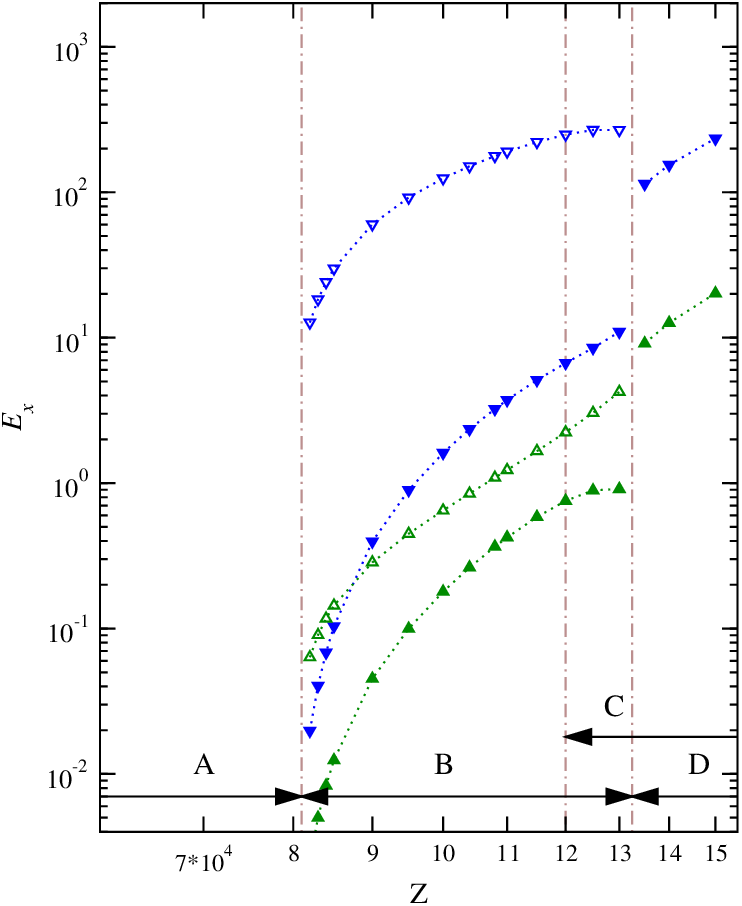,width=0.483\textwidth,clip=}
\end{center}
\caption{Time-averaged kinetic energy densities as functions of baroclinicity
$\Z$ for $\eta=0.3$,
  $n=2$, $N_\rho=2$, $\Pr=0.1$, $\tau=300$ and $\R=-5 \times 10^4$.
Full and empty symbols indicate equatorially-symmetric and -asymmetric
energy components, respectively. Black  circles, red squares, green
triangles-up and blue  triangles-down indicate the energy components
$\overline E_p^{s,a}$,  $\overline E_t^{s,a}$, $\widetilde E_p^{s,a}$,
$\widetilde E_t^{s,a}$, respectively. Axially-symmetric and
axially-asymmetric components are plotted in the left and the right
panels, respectively. Vertical dash-dotted lines 
indicate transition points. The ranges over which the four distinct
states are observed are indicated by arrows near the bottom
abscissa, with states \textbf{B} and \textbf{C} and \textbf{D} and
\textbf{C} co-existing as shown. Energy components not shown are at
least 10 orders of magnitude smaller than the ones shown. 
(Colour online)
}
\label{fig:020}
\begin{center}
\hspace*{-2.6mm}
\rule[-.3\baselineskip]{0pt}{70mm}
\epsfig{file=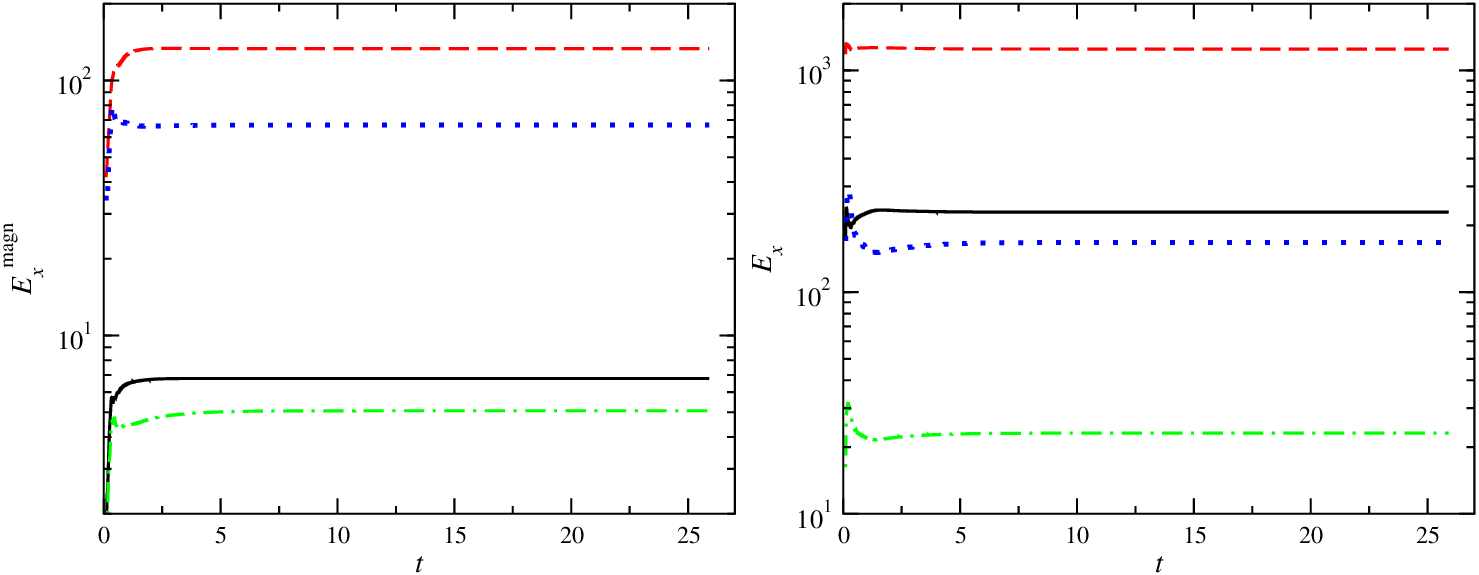,width=0.975\textwidth,clip=}
\end{center}
\caption{
Time series of magnetic dipolar energy densities (left panel) and
equatorially-symmetric kinetic energy densities (right pannel) in the
dynamo case $\Z = 17 \times 10^4$, $\Pm = 16$, for $\eta=0.3$, $n=2$,
$N_\rho=2$, $\Pr=0.1$, $\tau=300$  and $\R=-5 \times 10^4$.  Black
solid lines, red dashed lines, green dash-dotted lines and blue dotted
lines indicate the components $\overline E_p^{s}$,  $\overline
E_t^{s}$, $\widetilde E_p^{s}$, $\widetilde E_t^{s}$, and the
corresponding dipolar magnetic components, respectively.  All other
energy density components are smaller by more than 5 orders of
magnitude. (Colour online)
}
\label{fig:050}
\end{figure}

\section{Baroclinic flow instabilities}
%
The baroclinically-driven problem is invariant under a group of symmetry
operations including rotations about the polar axis
i.e.~invariance with respect to the coordinate transformation
$\varphi\to\varphi+\alpha$, reflections in the equatorial plane
$\theta \to \pi-\theta$, and translations in time $t \to t+a$, see
\citep{Gubbins1993}.
%
As baroclinicity $\Z$ is increased the available symmetries of the
solution are broken resulting in a sequence of states ranging from
simpler and more symmetric flow patterns to more complex flow patterns of lesser
symmetry as illustrated in figure 
\ref{fig:010}. 
In this respect the system resembles Rayleigh-Benard convection \citep{Busse2003}. 
The sequence starts with the basic axisymmetric, equatorially symmetric and
time-independent state with {a dominant wave number $k=1$ in the
radial direction} labelled \textbf{A} in figures \ref{fig:010} and 
\ref{fig:020}. 
An instability occurs at about {$\Z=8.1 \times 10^4$}
leading to a state  \textbf{B} characterized by a dominant azimuthal wavenumber
$m=2$ in the expansion of the solution in spherical harmonics $Y_l^m$,
i.e.~{while the full axisymmetry is broken,} a symmetry holds with
respect to the transformation
$\varphi\to\varphi+\pi$. {Simultaneously, the symmetry about  
the equatorial plane is also broken in state \textbf{B}.
At $\Z=13.25 \times 10^4$ a further transition to a pattern
labelled \textbf{D} occurs. While state \textbf{D} continues to
have a $m=2$ azimuthal symmetry, the symmetry about the equatorial
plane is now restored. In addition,  a dominant  radial wave number
$k=2$ develops as is evident, for instance, from the concentric two-roll
meridional circulation in state \textbf{D}. 
Remarkably, both states \textbf{B} and \textbf{D} coexist with a
steady pattern \textbf{C} that can be found for the range of values $\Z > 12
\times 10^4$. State \textbf{C} is axisymmetric, equatorially symmetric
and time-independent but differs from state \textbf{A} in that it
keeps a dominant radial wave number $k=2$. Which one of the coexisting
branches will be found in a given numerical simulation is determined
by the specific initial conditions used.
While all patterns in the sequence presented here remain
time-independent in their respective drifting frames of reference, we
expect that time-dependent solutions will be 
found for lower values of the Prandtl number since the Reynolds
number is increasing at a fixed baroclinicity $\Z$ and as a result
breaking of additional symmetries is likely to occur.}   

The bifurcations and their abrupt nature are also evident in
figure \ref{fig:020} where the time-averaged kinetic energy densities are
plotted with increasing baroclinicity $\Z$. 
At $\Z=0$ all energies vanish corresponding to the state of rigid body rotation 
with vanishing velocity field.  
In the range {$\Z<6 \times 10^4$} velocities grow linearly with $\Z$ and
the corresponding growth of kinetic energies is well described by the
empirical relations  
\begin{gather*}
\Emts =   5.647\times10^{-8}\; \Z^2, \qquad\qquad \Emps = 2.377\times 10^{-10}\; \Z^2,
\end{gather*}
indicating the dominance of toroidal motions as argued in earlier theoretical analyses
\citep{Busse1981,Busse1982,Zahn1992}. {While the particular
numerical factors in these expressions depend on the parameter values
used, the quadratic form of this dependence is expected to be generic.}
The abrupt jumps in energies at intermediate values of $\Z$ correspond
to the breaking of spatial symmetries. 
In state \textbf{A}, for instance, only the axially-symmetric
poloidal (i.e. meridional circulation) and toroidal (i.e. differential
rotation) kinetic energies have non negligible values. 
In state
\textbf{B} the non-axisymmetric components emerge with the
equatorially asymmetric ones being larger than the corresponding
equatorially symmetric ones. 
In state \textbf{D} equatorially asymmetric components become
negligible again
and in state \textbf{C} all non-axisymmetric energy components decay.
In the regions of hysteretic bistability $12\times10^4 < \Z <
13.25\times10^4$ and $\Z > 13.25\times10^4$ either of states
\textbf{B}  or \textbf{C} or  states \textbf{D} or \textbf{C} may be
realized, respectively,  depending on initial conditions.  

\section{Baroclinic dynamos}
Approximately 10\% of intermediate-mass and massive stars which are
mainly radiative are known to be magnetic \citep{Donati2009}. The most
popular hypothesis is that these are fossil fields remnants of an
early phase of the stellar evolution \citep{Donati2009}.   
The possibility of dynamos generated in stably stratified stellar
radiation regions has received only limited support in the literature
\citep[e.g.][]{Braithwaite2006}, because it is well known that dynamo action does not
exist in a spherical system in the absence of a radial component of
motion \citep{BullardGellman2954}. The latter is, indeed, rather weak
in the basic state \textbf{A} of low poloidal kinetic energy
{as} discussed above. However, with increasing baroclinicity
$\Z$, the  growing radial component of the velocity field strengthens
the probability of dynamo action.  

\begin{figure*}
\epsfig{file=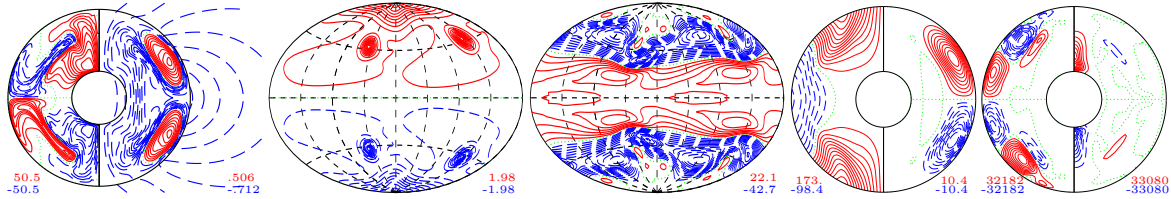,width=0.99\textwidth,clip=}
\caption{A dynamo solution at $\Z=17\times 10^4$ and $\Pm=16$ for 
$\eta=0.3$,
  $n=2$, $N_\rho=2$, $\Pr=0.1$, $\tau=300$ and $\R=-5 \times 10^4$. 
  {corresponding to the end of time in Figure  \ref{fig:050}}. 
The first plot shows isocontours of $\overline{B}_\varphi$
(left half) and meridional field lines $r \sin\theta\, \upartial_\theta
\bar{h} = $ const. (right half). The second plot
shows isocontours of radial magnetic field ${B}_r$ at
$r=r_o+0.1$. The third plot shows isocontours of radial velocity $u_r$
at $r=r_i+0.7$. The fourth plot shows  isocontours of
$\overline{u}_\varphi$ (left  half) and streamlines
$r\sin\theta(\upartial_\theta \overline{\pol})=$ const. (right half) in the meridional plane.
The last plot shows isocontours of azimuthally-averaged kinetic
helicity $\overline{\u{\bm \cdot}{\bm \nabla}\times\u}$ (left half) and magnetic helicity
$\overline{\B{\bm \cdot}{\bm \nabla}\times\B}$ (right half) in the meridional plane. 
{Minimal and maximal values and isocontour line types are denoted
following the convention of Figure \ref{fig:010}.} 
(Colour online)}
\label{fig:040}
\end{figure*}

To investigate this possibility the Lorentz force $(1/\rhobar)
({\bm \nabla}\times\B)\times\B$ has been added to the right-hand side of equation \eqref{govmod2.02} and the
equation of magnetic induction 
$$
\upartial_t \B = {\bm \nabla}\times(\u\times\B)+\Pm^{-1} \nabla^2 \B,
$$
has been added to equations \eqref{govmod2} to attain the full
magnetohydrodynamic form as used in the works of \cite{JonesBench} and
\cite{Simitev2015}. {Here, $\B$ is the magnetic flux density and
  $\Pm={\nu}/{\lambda}$ is the magnetic Prandtl number with $\lambda$
  being the magnetic diffusivity.} 
{After adding vanishingly small random magnetic fields as
initial conditions to equilibrated purely hydrodynamic solutions,
a number of solutions with growing  magnetic fields of dipolar
character have been obtained.}
As an example, figures \ref{fig:050} and \ref{fig:040}
demonstrate the magnetic field generated at {$\Z=17 \times 10^4$} and for the magnetic Prandtl
number {$\Pm=16$} which has been started from an equilibrated
neighbouring case to help convergence and reduce transients.
The dynamo solution quickly attains {a stationary state that we have
followed for nearly 2 magnetic diffusion time units, $t/\Pm$, as shown
in figure \ref{fig:050}}.
The ratio of poloidal to toroidal magnetic energy
is {$E_p^\text{magn}/E_t^\text{magn}=0.059$} and the ratio of magnetic to
kinetic total energy is {$E^\text{magn}/E^\text{kin}=0.127$}. 
{Furthermore, the energy density of the magnetic field
  $E^\text{magn}$ is comparable to the kinetic energy density of the
  poloidal component of the velocity field, $E_p^\text{kin}$.}
The magnetic field has a negligible quadrupolar component and a
large-scale dipolar topology {shown in figure \ref{fig:040} that does not} change in time,
resembling in this respect the surface fields of Ap-Bp stars \citep{Donati2009}.
{The surface structure of the magnetic field is characterised by two
prominent patches of opposite polarity situated at approximately
$45^\circ$ in latitude. The azimuthally-averaged toroidal field shows
a pair of hook-shaped toroidal flux tubes largely filling the outer
layer of the shell at lower latitudes as well as a pair of toroidal flux
tubes in the polar regions and parallel to the rotation axis. While
a large scale dipolar component emerges outside of the spherical
shell, the azimuthally-averaged poloidal field also shows a 
pair of octupole poloidal flux tubes confined to the interior of the
shell. Azimuthally-averaged kinetic and magnetic helicities are
plotted in the last column of figure \ref{fig:040}. These latter
quantities are important
in modelling the electromotive force in mean-field dynamo theories,
and in estimating the topological linkage of magnetic field lines, respectively.} 

\section{Conclusion}
In summary, we have described the bifurcations leading to
{non-axisymmetric and non-equatorially symmetric flow} states with
increasing baroclinicity. {The observed sequence of baroclinic
instabilities is rather different from the typical sequence of
convective instabilities familiar from the literature \citep[e.g.][]{SunSchubert1993,Simitev2003}.
Furthermore, w}e have demonstrated the possibility of
dynamo action in baroclinically-driven flows.  We remark that the
details of the bifurcation sequence depend on the choice of parameter
values. But preliminary additional computations indicate that the
general picture described here persists {when parameter values are
  changed by up to a factor of 10.} 
{While the dynamo solution discussed above is obtained for an
unphysically large value of the magnetic Prandtl number, w}e argue that magnetic fields in stably stratified stellar
interiors {may not necessarily be} of fossil origin as often
assumed and that dynamo action {may possibly} occur in the
radiative zones of rotating stars. Even in the solar interior below
the convection zone magnetic fields may be generated. This possibility
has not yet been investigated as far as we know.

\section*{Acknowledgments}
This work was supported by NASA [grant number NNX-09AJ85G] and the
Leverhulme Trust [grant number RPG-2012-600].


\end{document}